\documentclass[twocolumn,showpacs,preprintnumbers,amsmath,amssymb,epsf,superscriptaddress]{revtex4-1}

\usepackage{graphicx,epsfig}
\usepackage{dcolumn}
\usepackage{bm}

\begin{document}


\title{Propagation velocity of slip front and emergence of macroscopic static friction in the system with vanishing local static friction}
\author{Takehito Suzuki}
\email{t-suzuki@phys.aoyama.ac.jp}
\author{Hiroshi Matsukawa}
\affiliation{Department of Physics and Mathematics, Aoyama Gakuin University, 5-10-1 Fuchinobe, Chuo-ku, Sagamihara, Kanagawa 252-5258, Japan}


\begin{abstract}
We investigate the propagation of the slip front in the elastic body on the rigid substrate.
We first obtain the slip profile and the slip front velocity of the steady state by employing the local friction law with the quadratic form of the slip velocity and with vanishing static friction stress. 
The macroscopic static friction stress emerges spontaneously, which is expressed in terms of the parameter emerging in the friction law.
For the model with viscosity, the macroscopic static friction stress again emerges spontaneously.
The analytical treatment gives estimations for two slip front propagation velocities. 
They corresponds to two different boundary conditions, and one of them describes the framework employed here. 
Linear Marginal Stability Hypothesis based on the linearized equation of motion shows that two slip front propagation velocities exist in this system, both of which coincide with the analytical solutions noted above. 
These imply that the linearized friction law dominantly governs the slip front propagation behavior. 
Seismological implications are also given based on the analytical and numerical results.
\end{abstract}

\maketitle

\section{Introduction} \label{secI}
When we apply the loading stress to the solid object on the solid substrate in its tangential direction and increase the force slowly, the block slips as a whole only when the force exceeds the critical value, the macroscopic static friction force. 
Though this behavior has been recognized for a long time, the mechanism determining the macroscopic static friction force has not been clarified yet. 
Moreover, local slip prior to the macroscopic sliding, which is called precursors, occurs in some systems (e.g., \cite{Ben, Ots}). 
These previous studies imply that such slip has no stress singularity ahead of the front, so that it can be arrested by local small perturbation such as small increase of normal stress. 
From physical, industrial and seismological viewpoints, understanding the relationship between the precursors and the macroscopic slip initiation is required in order to clarify the mechanism determining the macroscopic static friction stress.

Precursors show peculiar behaviors in many aspects. 
The front propagation velocity sometimes exceeds the elastic wave velocity, but in other cases it is much smaller than the elastic wave velocity \cite{Lan1, Rad}. 
This behavior cannot be explained in terms of crack dynamics because the crack tip velocity (rupture velocity) cannot exceed the P wave velocity. 
The understanding of such difference has not been achieved. 
Additionally, the propagating length of the precursors which leads to macroscopic slips is also discussed \cite{Ots, Rad, Amu1, Bar}. 
The instability of the precursors at certain critical propagation length is investigated analytically \cite{Ots}. 
The effect of stress distribution on the propagating length of precursors is treated using constitutive laws consisting of viscoelasticity and the local Amontons' law with velocity-weakening friction \cite{Ots, Rad}. 
How the loading condition affects the propagating length of the precursor is also investigated \cite{Amu1}. 
Bar-Sinai $et \ al.$ \cite{Bar} employed a kind of rate-and-state dependent friction law, which leads to the friction law depending on the slip velocity nonlinearly in the steady state, and found that the velocity-strengthening part determines the critical nucleation size for the occurrence of macroscopic sliding.
Analytical understanding of the local friction law nonlinearly depending on the slip velocity, however, has not been clarified. 

Geophysical studies have also contributed to understanding behaviors of macroscopic slip initiation. 
Before the main shock, foreshocks are observed in many cases, which are examples of precursors.
Many aspects of foreshocks have been investigated in geophysical viewpoints. 
The hypocenters of foreshocks of the 2011 $M_\mathrm{w}$ 9.0 Tohoku-Oki Earthquake are widely known to migrate to the hypocenter of the main shock \cite{Kat}, but the mechanism of the behavior has not been clarified yet. 
Slow earthquakes, which have slip velocity and fault-tip-propagation velocity negligibly small compared to those of ordinary earthquakes, have also been observed \cite{Oba}.
The relation between slow earthquakes and ordinary earthquakes, in particular, whether the slow earthquakes induce ordinary earthquakes or not is now a controversial problem.

To understand behavior of macroscopic slip initiation, we consider slip front propagation (SFP) into an intact, homogeneous area under the condition of constant loading stress at one end of the system. 
If the steady SFP with finite propagation velocity is observed, the loading stress is regarded as exceeding the macroscopic static friction stress. 
If such a steady SFP does not emerge, the macroscopic slip is not considered to emerge. 

Spontaneous SFP has been understood by regarding the SFP as propagation of a front of a stable state into an unstable state for the governing equation, employing the Linear Marginal Stability Hypothesis (LMSH) \cite{Lan2, Mye, Sha}.
The LMSH has been employed widely to investigate the dynamics of fronts or domain walls propagating spontaneously into an unstable state described by nonlinear governing equation \cite{Saa1, Saa2}. 
It gives the solidification front speed \cite{Dee}, the chemical reaction front speed (e.g., \cite{Saa1, Saa2}) and the slip front velocity between block and substrate (e.g., \cite{Lan2, Mye}). 
This hypothesis asserts that even if the governing equations are nonlinear, suitably linearized model gives correct front behavior such as exact front propagating velocity. 
This hypothesis has actually been applied to the system with friction law linearized by the slip velocity \cite{Lan2, Mye}, while systematic and analytical 
treatment for various types of friction laws depending on the slip velocity has not been examined.

We study an elastic block on a rigid substrate and apply the stress acting on the left side in the direction tangential to the substrate surface. 
We first employ the local friction law depending on the quadratic form of the slip velocity without static friction stress. 
We analytically obtain the slip profile along the propagation direction and the front velocity of the steady state. 
The solutions of the slip and the slip velocity give the conditions with which macroscopic static friction stress emerges spontaneously. 
The applied stress on the left side is a controlling parameter, and the critical value of the controlling parameter is given in terms of the parameters in the friction law. 
We then introduce the viscosity of the block into the system and obtain two front propagation velocities analytically.
One of them is related with the boundary condition employed here and is smaller than that in the absence of the viscosity. 
The other one is found to correspond to another boundary condition.
We also obtain the macroscopic static friction stress.
We then employ LMSH and show that two front propagation velocities appear.
Both of them  are exactly the same as those obtained above.
Finally, we numerically investigate the SFP of visco-elastic systems and obtain macroscopic static friction and the propagation velocity.
They are not exactly the same as the analytic ones, but the analytic ones are found to give rough estimations of the macroscopic static friction stress and the propagation velocity.
Whether the macroscopic slip occurs or not, and how the front propagation is determined, are understood systematically in terms of the parameters of the friction law. 
We give some implications for precursors and slow earthquakes based on the slip with the driving stress less than the macroscopic static friction stress.

\section{MODEL WITH VELOCITY-DEPENDENT LOCAL FRICTION LAW} \label{secMWVD}

\subsection{Model and definition of slip front propagation} \label{secMod}

We consider a block on a rigid and fixed substrate and apply the stress acting on the left side in the direction tangential to the substrate surface (side-loading stress). 
The block is assumed to be one-dimensional (1D) system along $x$ direction and an infinite homogeneous medium. 
Deformation of the block is restricted to the $x$ direction.
The side-loading stress with certain strength is applied at $t \ge t_0$, where $t_0$ is a constant, along $x$ direction by pushing left side of the block. 
We consider the front propagation running from left to right. 
We thus require the boundary conditions $\lim_{x \to -\infty} \partial u/\partial x =p_{-\infty}(<0)$ and $\lim_{x \to \infty} \partial u/\partial x = 0$ for the slip propagation, where $u(x,t)$ is the slip displacement of the block at the position $x$ and the time $t$, and $p_{-\infty}$ is constant. 
Note that the constant $p_{-\infty}$ represents both constant stress and strain at the left side of the block because the Young modulus $E_1$ will be assumed to be constant in this study.
We consider here Slip Front Propagation (SFP) into an intact area, i.e., the system is homogeneous and $u(x,t)=0$ before SFP arrives under the boundary condition noted above. 


It is to be noted that we discuss friction stress, not friction force, in this paper. 
Since the model is infinite, the normal force acting on the substrate and the friction force diverge. 
We should adopt a stress, not a force, as a controlling parameter for discussion independent of system size.

We then derive nondimensionalized equation of motion. The equation with dimensions is given by
\begin{equation}
\rho \frac{\partial^2 u}{\partial t^2} =E_1 \frac{\partial^2 u}{\partial x^2}-\tau_{\mathrm{fric}},
\end{equation}
where $\rho$ is the mass density of the block, and $\tau_{\mathrm{fric}}$ is the local friction stress. Note that the dimension of $\tau_{\mathrm{fric}}$ is [Pa/m] here. If we use the characteristic length $L_0$ and time $T_0$, we have
\begin{equation}
\frac{\partial^2 \tilde{u}}{\partial \tilde{t}^2} =\frac{E_1 T_0^2}{\rho L_0^2} \frac{\partial^2 \tilde{u}}{\partial \tilde{x}^2}-\frac{T_0^2}{\rho L_0} \tau_{\mathrm{fric}},
\end{equation}
where $\tilde{u}, \tilde{x}$ and $\tilde{t}$ are normalized slip displacement, space and time, respectively. We cannot determine $L_0$ and $T_0$ uniquely because the only single independent parameter with dimensions, $E_1/\rho$, exists in the present framework. Because the friction stress $\tau_{\mathrm{fric}}$ can take various forms, we do not use the coefficients emerging in $\tau_{\mathrm{fric}}$ to normalize the governing equations. Hence, we put $E_1 T_0^2/\rho L_0^2 \equiv 1$; note that the elastic wave velocity is given by unity below. We will investigate the behavior of $\tilde{u}$ in terms of $\tilde{x}$ and $\tilde{t}$. Henceforth, we will describe these normalized values as $u$, $x$ and $t$, respectively. The boundary conditions  $\lim_{x \to -\infty} \partial u/\partial x =p_{-\infty}(<0)$ and $\lim_{x \to \infty} \partial u/\partial x = 0$ do not change with this change of the notation. Additionally, the equation of motion is expressed as
\begin{equation}
\ddot{u} = u'' - \tau, \label{eqeom1}
\end{equation}
where the dot and prime represent the differentiations with respect to time and space, respectively, and $\tau \equiv T_0^2 \tau_{\mathrm{fric}}/\rho L_0$ is the normalized local friction stress.

\begin{figure}[tbp]
\centering
\includegraphics[width=7.5cm]{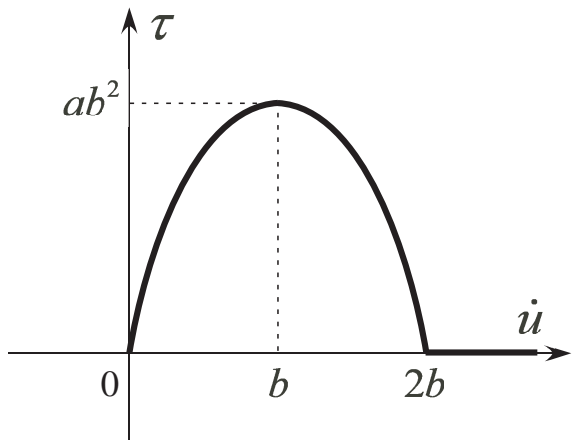}
\caption{Constitutive law of the local friction stress as a function of  the slip velocity.}
\label{FigCL}
\end{figure}

We first assume that $\tau$ takes the form of
\begin{equation}
\tau = a \dot{u} (2b -\dot{u}) [H(\dot{u})-H(\dot{u}-2b)], \label{eqtau}
\end{equation}
where $a$ and $b$ are positive constants, $H(\cdot)$ is the Heaviside function (see also Fig. $\ref{FigCL}$). With this local friction law, the friction stress changes from the velocity-strengthening to the velocity-weakening behaviors at $\dot{u}=b$ with increasing slip velocity, and vanishes for $\dot{u} \ge 0$. 
This assumption enables us to treat the problem analytically, because the friction stress becomes a single valued function of $\dot{u}$. 
It should also be noted that for large $\dot{u}$ regime, $\dot{u} >2b$, the friction stress vanishes. 
Actually, in some seismological systems owing to, e.g., the thermal pressurization \cite{Suz06} or melting at the sliding plane \cite{Hir}, sliding friction stress almost vanishes in large velocity regime.

The slip front is defined to be located where the friction stress becomes less than a certain critical value after exceeding the maximum value $ab^2$, i.e., $\dot{u} \le 2b$ and $\dot{u} \sim 2b$, based on the analogy of dynamic crack tip propagation (the critical value does not affect the following study if it is negligibly small to $ab^2$). 
For the dynamic crack tip propagation, the crack tip is considered to pass when the friction stress achieves the dynamic friction stress. 
Note that the driving force acts in the whole system, not one side of the system, for the dynamic crack tip propagation, so that the model is not exactly the same as the present one.
If the slip velocity over a whole system decays with increasing time, we consider that SFP vanishes.

\subsection{Exact solution for slip profile} \label{secExact1}

We first obtain the slip profile with SFP. 
From Eqs. ($\ref{eqeom1}$) and ($\ref{eqtau}$), we have 
\begin{equation}
\ddot{u} = u'' + (a \dot{u}^2 - 2ab \dot{u}) [H(\dot{u})-H(\dot{u}-2b)].  \label{eqGov}
\end{equation}
We consider here the steady state solution in Eq. ($\ref{eqGov}$) propagating with the constant SFP velocity $v$, which means $u(x,t)$ has a form $u(x_1=x-vt)$.

We first assume $0 < \dot{u} < 2b$ over a whole space. We can obtain the analytical solution for the steady state, which is expressed as
\begin{equation}
u(x_1) = \frac{2b}{\gamma v} \ln \left( 1 - \frac{vp_0}{2b+vp_0} e^{-\gamma x_1} \right), \label{eqan-u}
\end{equation}
and its temporal derivative
\begin{equation}
\dot{u}(x_1)=\frac{2bvp_0 e^{-\gamma x_1}}{vp_0 e^{-\gamma x_1} -2b-vp_0}, \label{eqSolUu}
\end{equation}
where $\gamma=2abv/(1-v^2)$ and $p_0=p(x_1=0) <0$. Note that $\gamma >0$ is required for the solution expressed by Eqs. (\ref{eqan-u}) and (\ref{eqSolUu}), which leads to $v < 1$, i.e., $|p_{-\infty}|>2b$ (see details in Appendix \ref{secAA}).
We can confirm that $\lim_{x_1 \to -\infty} \dot{u}(x_1)=2b$ and $\lim_{x_1 \to \infty} \dot{u}(x_1)=0$.

The solution (\ref{eqan-u}), (\ref{eqSolUu}) is, however, unstable. To show this, we put $u_1=u+\delta u$ and linearize Eq. (\ref{eqGov}) with respect to $\delta u$ in the region $0<\dot{u}<2b$. This reads
\begin{equation}
\delta \ddot{u} = \delta u'' +2a(\dot{u}-b) \delta \dot{u} [H(\dot{u})-H(\dot{u}-2b)],
\end{equation}
which is a wave equation with the velocity-weakening or -strengthening behavior depending on the sign of $2a(\dot{u}-b)$ with the condition $H(\dot{u})-H(\dot{u}-2b)=1$; if $2a(\dot{u}-b)$ is positive, the velocity-weakening is described and the system is unstable, whereas if it is negative, the velocity-strengthening is represented and the system is stable. We can conclude that the solution (\ref{eqan-u}), (\ref{eqSolUu}) is unstable because it includes the region where $\dot{u}-b > 0$. 

We now find the stable solution for Eq. (\ref{eqGov}) with the assumption $|p_{-\infty}|>2b$. Actually, we can easily confirm that
\begin{equation}
\dot{u} (x_1=x-1 \times t)=|p_{-\infty}| H (t-x-C), \label{eqSolUs}
\end{equation}
where $C$ is a constant, is a solution for Eq. (\ref{eqGov}). We can also see that this solution is stable because the friction stress term vanishes [$H(\dot{u})-H(\dot{u}-2b)=0$] and Eq. (\ref{eqGov}) reduces to the wave equation $\ddot{u}=u''$. This solution is stable and physically realized. The steady SFP is observed and the block is considered to slip macroscopically. Furthermore, we can clearly confirm that the SFP velocity is given by unity, i.e., the elastic wave velocity.

When $|p_{-\infty}| < 2b$, the solution (\ref{eqan-u}), (\ref{eqSolUu}) does not exist. Additionally, the solution (\ref{eqSolUs}) does not emerge in this case. If we assume $|p_{-\infty}| < 2b$ for Eq. (\ref{eqSolUs}), the friction stress term in Eq. (\ref{eqGov}) does not vanish, whereas $\ddot{u}=u''$. Hence, Eq. (\ref{eqSolUs}) is not a solution of Eq. (\ref{eqGov}). Nonetheless, the steady stable state exists. We here consider the state below:
\begin{equation}
u(x_1=x-0 \times t)=p_{-\infty} x+C', \label{eqSolUs2}
\end{equation}
where $C'$ is a constant.
The state approaching (\ref{eqSolUs2}) in the limit $t \to \infty$ with infinitely small slip velocity is a stable solution for Eq. (\ref{eqGov}). 
Note that the state (\ref{eqSolUs2}) realizes the constant strain $p_{-\infty}$ over the whole plane.
Therefore, Eq. (\ref{eqSolUs2}) satisfies $\lim_{x_1 \to -\infty} \partial u/\partial x_1 =p_{-\infty}$, which is one of the boundary conditions adopted here, and the state approaching (\ref{eqSolUs2}) also satisfies the condition.
However, the state (\ref{eqSolUs2}) does not satisfy $\lim_{x_1 \to \infty} \partial u/\partial x_1 =0$, which is the other boundary condition.
Nonetheless, the state approaching (\ref{eqSolUs2}) satisfies this boundary condition within finite time.
In addition, the state (\ref{eqSolUs2}) shows the zero propagation velocity.
When $|p_{-\infty}| < 2b$, the steady stable state approaching (\ref{eqSolUs2}) emerge and the steady SFP cannot be observed.

Actually, even when $|p_{-\infty}|>2b$, the solution  (\ref{eqSolUs2}) mathematically exists. However, the loading condition here (the side-loading stress with certain strength is applied at $t \ge t_0$ along $x$ direction) does not generate such a state, since infinitely large slip velocity at the side-loading point and its relaxation to $|p_{-\infty}|$ are expected, indicating that the slip velocity profile approaches Eq. (\ref{eqSolUs}).

We can understand whether the steady SFP emerges in terms of  $p_{-\infty}$ based on the conclusion here. 
For the case of $|p_{-\infty}| > 2b$, the steady SFP appears, and this slip with finite velocity propagates into the whole system. 
On the other hand, if $|p_{-\infty}| < 2b$, $\dot{u}$ approaches zero over the whole plane with increasing time, indicating no macroscopic slip. 
These statements imply that the macroscopic static friction stress appears spontaneously, even though  it does not exist in the local friction law. 
The critical side-loading stress at $x \to -\infty$ is given by $2b$, which is the lower limit of the side-loading strain generating the steady SFP.
Since the Young modulus is a constant, we can regard the lower limit as the macroscopic static friction stress in the normalized system. 
If the loading stress is smaller than this critical value, no macroscopic slip appears, whereas the slip diverges with $t \to \infty$. 
This behavior may be related with creep motion because such slip has negligibly small velocity.

The emergence of the steady SFP and the macroscopic static friction stress has been understood based on Eqs. (\ref{eqSolUs}) and (\ref{eqSolUs2}). 
We will show below that such emergence is not unique to the friction law depending on the slip velocity with the quadratic form as shown in Eq. (\ref{eqtau}).
Details of the friction law do not affect the existence of the steady SFP and the macroscopic static friction stress. 

Let us assume that the friction stress is given by
\begin{equation}
\tau=-\dot{u} g(\dot{u})(\alpha_1 \dot{u} - \alpha_2)[H(\dot{u})-H(\dot{u}-\alpha_2/\alpha_1)], \label{eqtau2}
\end{equation}
where $\alpha_1$ and $\alpha_2$ are positive constants and $g(\dot{u})$ is an arbitrary continuous function satisfying $g(\dot{u})>0$ and $|\partial g(\dot{u})/\partial \dot{u}| < \infty$ for $0<\dot{u}<\alpha_2/\alpha_1$. With this system, we first derive the boundary value of $\dot{u}$ for the steady state. To derive it, we rewrite Eq. (\ref{eqeom1}) in terms of the friction stress (\ref{eqtau2}) and $p=p(x_1=x-vt)$ to have
\begin{equation}
(v^2-1)\frac{dp}{dx_1} =g(-vp) vp(v \alpha_1 p+\alpha_2),
\end{equation}
which leads to
\begin{equation}
\int_{p_{-\infty}}^{p(x_1)} \frac{dp}{g(-vp) p(v \alpha_1 p +\alpha_2)} = -\int_{-\infty}^{x_1} \frac{vdx'_1}{1-v^2}. \label{eqIntdpdx}
\end{equation}
Since the integrand in the left hand side of Eq. (\ref{eqIntdpdx}) diverges at $p=0$ and $-\alpha_2 /v \alpha_1$, the integral is meaningful in the range $-\alpha_2/v \alpha_1 < p <0$. In addition, because the integral of the right hand side clearly diverges, $p_{-\infty}=-\alpha_2/v \alpha_1$ must be satisfied. 
We can conclude that $\dot{u}|_{x_1 \to -\infty}=\partial u/ \partial t|_{x_1 \to -\infty}=-v \partial u/ \partial x|_{x_1 \to -\infty} =-v p_{-\infty} =\alpha_2/ \alpha_1$ is the boundary value of $\dot{u}$ required for the steady SFP.

We then linearize Eq. (\ref{eqeom1}) using (\ref{eqtau2}) and $u_1 =u+ \delta u$ to have
\begin{equation}
\delta \ddot{u} = \delta u'' +
[g(\dot{u}) (2 \alpha_1 \dot{u} -\alpha_2) +\dot{u} \frac{\partial g(\dot{u})}{\partial \dot{u}} (\alpha_1 \dot{u} -\alpha_2)] \delta \dot{u}.
\end{equation}
In the region where $\dot{u} \sim \alpha_2/\alpha_1$, the term $\dot{u} \partial g(\dot{u})/\partial \dot{u} (\alpha_1 \dot{u} -\alpha_2)$ is negligible, whereas the term $g(\dot{u}) (2 \alpha_1 \dot{u} -\alpha_2)$ is nonnegligible and positive. We can conclude that there must exist the region where the solution is unstable, and that the stable steady state does not emerge. This conclusion is consistent with the friction law of the quadratic form of the slip velocity. Moreover, the solutions (\ref{eqSolUs}) and (\ref{eqSolUs2}) are the solutions also for Eq. (\ref{eqeom1}) with the friction law (\ref{eqtau2}). Hence,  we can conclude that the criticality about emergence of the static friction stress is a universal phenomenon and it is not unique to the friction law with the quadratic form of $\dot{u}$. 
Only the assumption required for the criticality is that the friction stress becomes zero with $\dot{u}$ larger than a certain value; e.g., for the friction law (\ref{eqtau2}), the friction stress vanishes where $\dot{u} \ge \alpha_2/\alpha_1$. We can also interpret that $\alpha_2/\alpha_1$ gives the macroscopic static friction stress.

\subsection{Numerical calculations} \label{secNum1}

\begin{figure}[tbp]
\centering
\includegraphics[width=8.5cm]{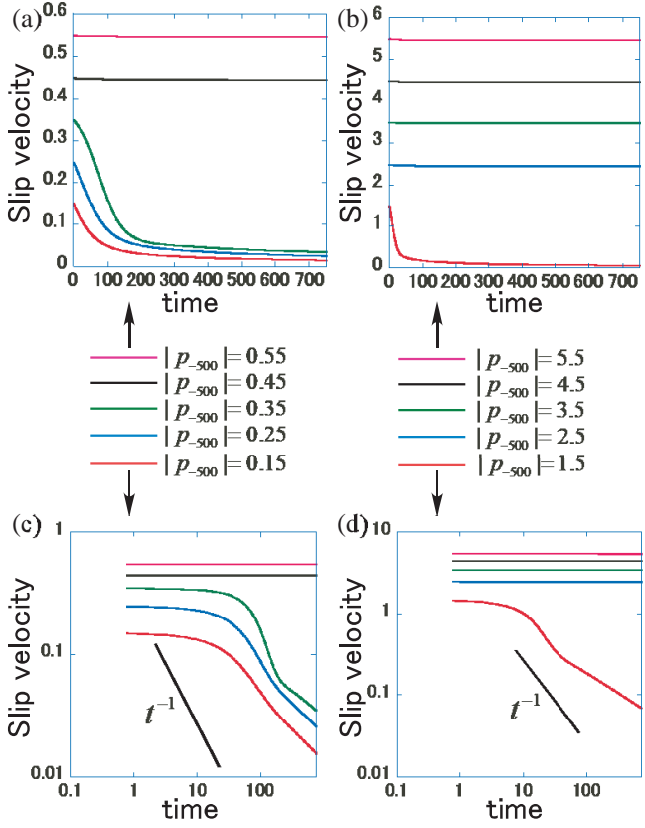}
\caption{Temporal changes in the slip velocity at the side-loading point $x=-500$. 
The critical value $2b$ is (a, c) $0.4$ and (b, d) $2$, and the cases (a, c) $|p_{-500}|=0.15, 0.25, 0.35, 0.45$ and $0.55$ and (b, d) $|p_{-500}|=1.5, 2.5, 3.5, 4.5$ and $5.5$ are calculated. (a, b) Linear plot, and (c, d) log-log plot.}
\label{FigDec}
\end{figure}

By solving numerically the equation of motion ($\ref{eqGov}$) with the Runge-Kutta method of the fourth-order accuracy, we can confirm the analytical results obtained in Sec. \ref{secExact1} (Fig. $\ref{FigDec}$). 
The elastic medium spans from $x=-500$ to $500$ and the side-loading point is  at $x=-500$, so that $p_{-\infty}$ appearing in the discussion above should be regarded as $p(-500,t) \equiv p_{-500}$. 
The viscosity term $\eta_{\mathrm{num}} \dot{u}''$ is introduced into the numerical calculations to ensure the numerical stability. 
The viscosity $\eta_{\mathrm{num}}$ is fixed to be $10^{-2}$ in the following results, but the small change of  this value does not affect the results. 
We tested two sets of the parameters $(a,b)=(0.1, 0.2)$ and $(0.1,1)$, which results in $2b=0.4$ and $2$, respectively. 
The systems of  five cases for the values of $p_{-500}$ are calculated for each set.

\begin{figure}[tbp]
\centering
\includegraphics[width=8.5cm]{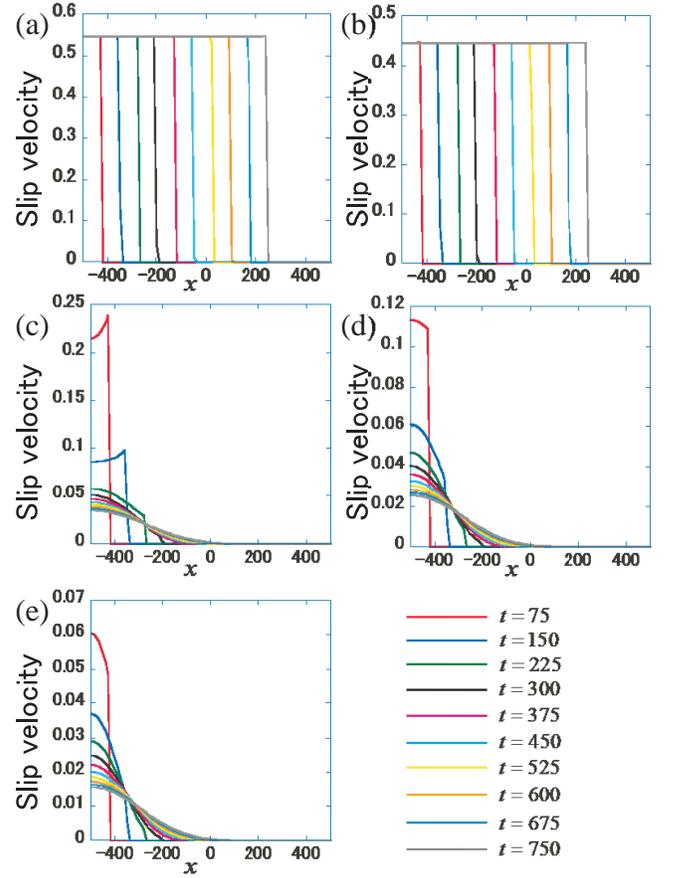}
\caption{Spatiotemporal changes in the slip velocity profile. 
The critical value $2b$ is $0.4$, and the five cases $|p_{-500}|=$ (a) $0.55$, (b) $0.45$, (c) $0.35$, (d) $0.25$, and (e) $0.15$ are calculated. 
Difference in line colors describes that of the time. 
The SFP velocity obtained analytically is equal to unity, and it can be confirmed from (a) and (b). 
For example, the slip front for the black line ($t=300$) exists at $x=-200=-500+300=-500+300 \times 1$.}
\label{FigSTV}
\end{figure}

We can confirm that the slip velocity at the loading point approaches $|p_{-500}|$ for the case of $|p_{-500}| > 2b$ from Fig. $\ref{FigDec}$. 
Figures $\ref{FigSTV}$(a, b) clearly show that the constant slip velocity $|p_{-500}|$ is observed where the slip front has passed, and the steady SFP appears with the propagation velocity unity.
In Fig. \ref{FigSTV}, the spatiotemporal profile of the slip velocity is shown only for the set $(a, b)=(0.1,0.2)$ because the behavior is qualitatively same for the set $(a, b)=(0.1,1)$.


For the case of $|p_{-500}| < 2b$, the slip velocity  at the side-loading point decays monotonically with increasing time  as predicted from the analytical treatment, which concludes that the steady SFP does not exist. 
Note that the velocity decays more slowly than $t^{-1}$ as shown in Figs. $\ref{FigDec}$(c, d), and the cumulative slip at the side-loading point does not converge. 
This is consistent with the statements in Sec. \ref{secExact1} because the strain tends to approach $p_{-500}$ over the whole plane; in the numerical calculations, the system is finite and the uniform strain can be achieved. 
However, the strain field is not shown because achieving the uniform strain field takes so long time that we can not show the state in Figs. $\ref{FigDec}$ and $\ref{FigSTV}$ due to the limiting computational time. 
We can also confirm from Figs. $\ref{FigSTV}$(c-e) that the steady SFP does not emerge in the case of $|p_{-\infty}|<2b$.

The results numerically obtained are consistent with those analytically obtained. The emergence of the static friction stress has also been confirmed.

\section{MODEL WITH VISCOSITY AND VELOCITY-DEPENDENT FRICTION LAW} \label{secMWVVD}

\subsection{Analytical study} \label{subAna}

\subsubsection{Exact results for the SFP velocity} \label{secExact2}

The front velocity of the steady SFP is also obtained analytically in the visco-elastic system. 
The governing equation is given by
\begin{equation}
\ddot{u}= u'' + \eta \dot{u}'' + (a \dot{u}^2 -2ab \dot{u}) [H(\dot{u})-H(\dot{u}-2b)], \label{eqeom2}
\end{equation}
where $\eta$ is the viscosity of the block. 
We consider the solution depending on $x_2 \equiv -x+vt$, where $v$ is the SFP velocity. 
With this definition, Eq. ($\ref{eqeom2}$) leads to
\begin{equation}
(1 - v^2) u'' + v \eta u''' + av^2 u'^2 -2abv u' =0, \label{eqSS1}
\end{equation}
where the prime denotes the differentiation with respect to $x_2$. 
We assume $0<\dot{u}<2b$. 
Equation ($\ref{eqSS1}$) is expressed by using $P=u'$ as
\begin{equation}
P'' = -\frac{1 - v^2}{\eta v} P' -\frac{a}{\eta} ( v P^2 -2b P ). \label{eqp2}
\end{equation}
By normalizing  Eq. ($\ref{eqp2}$), we get
\begin{equation}
\frac{\partial^2 q}{\partial X^2} = -\mu \frac{\partial q}{\partial X} -\frac{\partial}{\partial q} \left( \frac{q^2}{2} - \frac{q^3}{3} \right), \label{eqq2}
\end{equation}
where $q(x_2)= -(v/2b) \left[ P(x_2) -2b/v \right], \ X = x_2 \sqrt{2ab/\eta}, \ \mu=(1-v^2)/(\sqrt{2ab \eta} v )$.

In Eq. ($\ref{eqq2}$) we can regard $q$ and $X$ as the displacement and time, respectively.
Then Eq. ($\ref{eqq2}$)  is the equation of motion for the particle under the potential $U=q^2/2 - q^3/3$ with damping force proportional to the slip velocity, $-\mu \partial q/\partial X$.
Here, $\mu$ corresponds to the damping constant. 
It is clear that $U$ has a stable point at $q=0$ and an unstable point at $q=1$. 
We consider a solution propagating from the unstable state ($q(-\infty)=1$) to the stable state ($q(\infty)=0$). 
It should be emphasized that $\mu$ is a function of $v$ and has the critical value corresponding to the critical damping above which the solution approaches $q=0$ monotonically. 
In Eq. ($\ref{eqq2}$), the critical value is $\mu=2$ as easily shown. 
Aronson and Weinberger mathematically showed that the spontaneous front propagation chooses the critical value \cite{Aro}. 
We then obtain $(1 - v^2)/(\sqrt{2ab \eta} v) = 2$, which concludes $v=\sqrt{1 + 2ab \eta} -\sqrt{2ab \eta} \equiv v_{c-}$.
Here we have chosen only the positive velocity solutions, and the velocity $v_{c-}$ satisfies $\mu >0$, which must be satisfied for the existence of the steady state \cite{Aro}. 
This is the analytical solution of the steady SFP velocity with Eq. ($\ref{eqeom2}$).

For the boundary values $q(-\infty)=1$ and $q(\infty)=0$, we have $P(-\infty)=0$ and $P(\infty)=2b/v_{c-}$.
In order to understand physical meaning of these boundary values, we also introduce $p(x_3) \equiv \partial u/\partial x_3$, where $x_3 =x-vt$. 
The field $p(x_3)$ is a real strain field, and we have $p|_{x_3 \to -\infty}=-2b/v_{c-}$ and $p|_{x_3 \to \infty}=0$. 
This steady state is realized with zero strain over the whole system in the initial state and by letting $p|_{x_3 \to -\infty}=-2b/v_{c-}$ after $t=t_0$. 
Additionally, using the relationship $\dot{u}=\partial u/\partial t =-v \partial p/\partial x$, we have $\dot{u}|_{x_3 \to -\infty}=2b$ and $\dot{u}|_{x_3 \to \infty}=0$. 
We can interpret that the initial state is the zero slip velocity state, and setting the slip velocity as $2b$ at the left side of the block at $t=t_0$ will generate the steady state after $t \to \infty$. 
This corresponds to SFP to the right direction with constant propagation velocity $v_{c-}$.

We have another solution if we put $x_2=-x-vt$, which concludes $v=\sqrt{1 + 2ab \eta} +\sqrt{2ab \eta} \equiv v_{c+}$. With $x_2=-x-vt$, we employ $x_3=x+vt$ and have $p|_{x_3 \to -\infty}=2b/v_{c+}$ and $p|_{x_3 \to \infty}=0$. 
This steady state is realized by fixing the strain to be $2b/v_{c+}$ over the whole system in the initial state and letting $p|_{x_3 \to \infty}=0$ after $t=t_0$. 
The slip front propagates in a left direction with the constant speed $v_{c+}$. 
In terms of the slip velocity, the initial state is that where $\dot{u}=2b$ over the whole system.
If we make the slip velocity zero at the right side of the block at $t=t_0$, the steady state will be generated after $t \to \infty$.  
Though this case does not coincide with the situation here, the velocity $v_{c+}$ actually has  physical meaning.

\subsubsection{Linear Marginal Stability Hypothesis} \label{secLMSH}

We extend the results obtained in the previous section to arbitrary friction laws here. 
For the purpose, we introduce Linear Marginal Stability Hypothesis (LMSH). 
First, we will introduce two forms of the fronts, ``extruding front'' and ``intruding front'', and show that the extruding and intruding front velocities, $v^{\mathrm{ex}}$ and $v^{\mathrm{in}}$, respectively, are exactly the same as $v_{c-}$ and $v_{c+}$, respectively, by linearizing the friction stress (\ref{eqtau}). 
Next, the hypothesis will indicate that only linearized form of the friction law plays an important role and the other details of the law does not affect the SFP velocities.

As mentioned in Sec. \ref{secI}, LMSH has been adopted widely to investigate the dynamics of fronts or domain walls propagating spontaneously into an unstable state \cite{Saa1}; the chemical reaction front speed (e.g., \cite{Saa1, Saa2}) and the SFP velocity between blocks and substrates (e.g., \cite{Lan2, Mye}) are examples of application of LMSH. 
However, previous studies about SFP \cite{Lan2, Mye} did not obtain the analytically exact SFP velocity, but only approximated form of the velocity was shown. 
This hypothesis requires linearizing the governing equations, the plane wave approximation of the solution around the propagating front, and two conditions about growth of disturbance and stability of propagation. It states that the characteristic frequency, wave number and propagating velocity of fronts can be derived by these requirements.

We here explain procedure for applying LMSH. We first define $s$ as a nondimensional variable characterizing the state of the system, like normalized slip distance or slip velocity for the motion of the continuum, and consider the dynamics of the spontaneous propagation of $s$. The system is assumed to be 1D. We treat the solution front intruding the unstable region. We consider two cases: one where $s=0$ is stable and this region intrudes the unstable region with $s \neq 0$, and the other one where $s=0$ is unstable and the stable region with $s \neq 0$ intrudes into this unstable region. The front for the former solution is called the extruding front, and that for the latter one is called intruding front in this paper. The front is mathematically defined to be located where only the terms $O(|s|)$ play important roles and the terms $O(|s|^2)$ becomes negligible in the governing equation. Consistency of this definition with previous treatment actually exists, as shown later in this section. 

\begin{figure}[tbp]
\centering
\includegraphics[width=8.5cm]{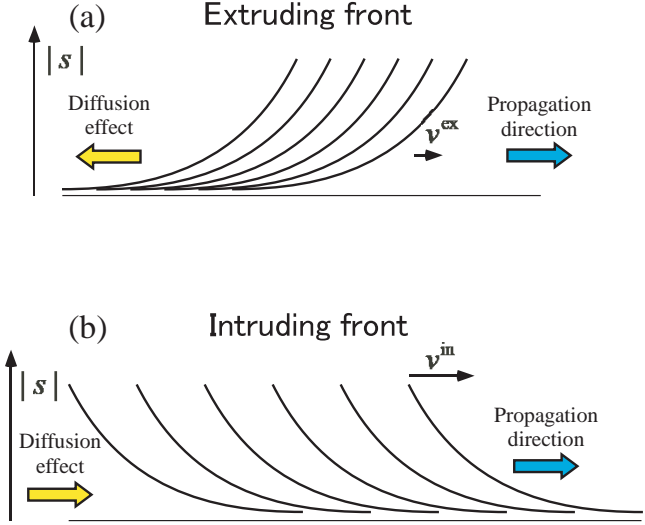}
\caption{Schematic representations of (a) the extruding front and (b) the intruding front in terms of $|s|$.}
\label{FigSch}
\end{figure}

For the front of $s$, we assume the plane wave  $s \sim \exp(\mp i(kx-\omega t))$ whose frequency $\omega$ and wave number $k$ are complex. This description results in $|s|=\exp [\pm (k_i x - \omega_i t)]$, where $k_i$ and $\omega_i$ are the imaginary parts of $k$ and $\omega$, respectively, and assumed to be nonnegative. Note that the front $\exp(k_i x -\omega_i t)$ and $\exp[-(k_i x -\omega_i t)]$ describes the extruding and intruding fronts, respectively, based on their definition (Fig. \ref{FigSch}; diffusion effect will be discussed later). We have four unknown parameters: $k_i$, $\omega_i$, $k_r$ being the real part of $k$, and $\omega_r$ being the real part of $\omega$. The parameters $k_r$ and $\omega_r$ are also assumed to be nonnegative. The four parameters can be determined from the viewpoint of LMSH by four independent equations: the real and imaginary parts of the dispersion relation, the growth stability and the propagating stability. The equations describing those stabilities are given by
\begin{equation}
\frac{\partial \omega_i}{\partial k_r}=0,  \label{eqGI1}
\end{equation}
\begin{equation}
\frac{\omega_i}{k_i} = \frac{\partial \omega_i}{\partial k_i}=c, \label{eqPI1}
\end{equation}
\begin{equation}
\frac{\partial \omega_r}{\partial k_i} =0, \label{eqGI2}
\end{equation}
\begin{equation}
\frac{\partial \omega_r}{\partial k_r} =\frac{\omega_r}{k_r} = c, \label{eqPI2}
\end{equation}
where $c$ is a spontaneous front propagation velocity (see details in Appendix \ref{secAB}). 

We apply LMSH for the frictional phenomenon between the substrate and the block with the viscosity assuming the friction law (\ref{eqtau}). In particular, we will show $v^{\mathrm{ex}}=v_{c-}$ and $v^{\mathrm{in}}=v_{c+}$ introduced in the previous subsection based on LMSH. We first clarify $v^{\mathrm{ex}}<v^{\mathrm{in}}$ in terms of physical implications. We expand the friction law appearing in equation of motion ($\ref{eqeom2}$) near the point where the friction stress vanishes, i.e., $\dot{u}=2b$, and assume $\dot{u}-2b<0$. This is because the region where $\dot{u} \sim 2b$ and $\dot{u}<2b$ is unstable, as mentioned in Sec. \ref{secExact1}, and the unstable region is overtaken by the region $\dot{u}=2b$. The slip front exists there, and we have a linear equation
\begin{equation}
\ddot{u}= u'' + \eta \dot{u}'' + 2ab (\dot{u} - 2b). \label{eqN1}
\end{equation}
This linearization is equivalent to neglecting $O(|\dot{u} - 2b|^2)$, which is consistent with the definition of the front in this subsection. Using $w=u-2bt$, Eq. (\ref{eqN1}) reduces to
\begin{equation}
\ddot{w}= w'' + \eta \dot{w}'' + 2ab \dot{w}. \label{eqeom3}
\end{equation}
Here we put $s \equiv \dot{w}=\dot{u}-2b$. Hence the linearization employed in Eq. (\ref{eqN1}) is equivalent to the linearization around $s=0$ noted above. We obtain
\begin{equation}
\ddot{s}= s'' + \eta \dot{s}'' + 2ab \dot{s}. \label{eqeom4}
\end{equation}

It should be emphasized that $s=\dot{w}$ obeys diffusion equation ($\dot{s}=\eta s''$) if we neglect the first and third terms of the right hand side of Eq.  ($\ref{eqeom3}$), and $s$ obeys wave equation ($\ddot{s}=s''$) if we neglect the second and third terms of the right hand side of Eq. ($\ref{eqeom4}$). Therefore, not only the slip front propagates as the wave equation, but also acceleration or deceleration can occur by the diffusion. 
Let us consider the amplitude of the front, $|s|=\exp [\pm (k_i x - \omega_i t)]$, at a certain point $x$. The diffusion effect enhances the amplitude $|s|$ because the second derivative of $\exp[\pm(k_i x -\omega_i t)]$ with respect to $x$ is always positive. Furthermore, $|s|$ decreases (increases) after the extruding front $\exp(k_i x -\omega_i t)$ (intruding front $\exp[-(k_i x -\omega_i t)]$) passes because $\exp(k_i x -\omega_i t)$ ($\exp[-(k_i x -\omega_i t)]$) is an increasing (decreasing) function in terms of $k_i x-\omega_i t$. We can conclude that the diffusion effect suppresses (enhances) the propagation for the extruding (intruding) front (Fig. \ref{FigSch}), and that $v^{\mathrm{ex}}<v^{\mathrm{in}}$. In addition, the front propagation velocity must be the elastic wave velocity in the absence of the diffusion effect, so that we have $v^{\mathrm{ex}}<v_e<v^{\mathrm{in}}$, where $v_e$ is the elastic wave velocity (unity here).

\begin{figure}[tbp]
\centering
\includegraphics[width=8.5cm]{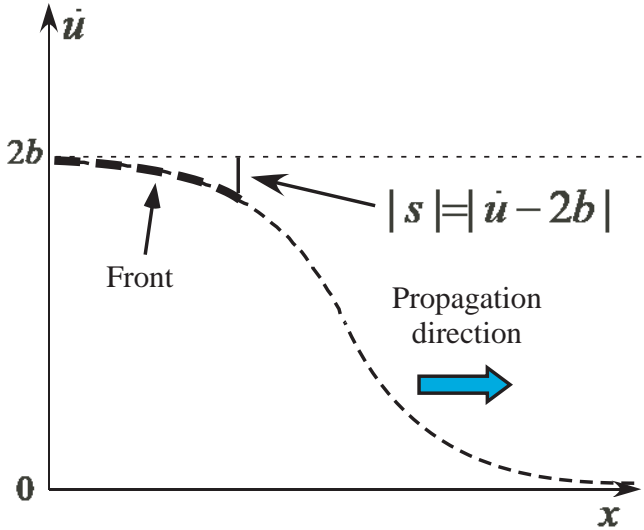}
\caption{Schematic representation of the SFP in terms of the slip velocity. Compare the form of the slip front with Fig. \ref{FigSch}(a).}
\label{FigFr}
\end{figure}

Note that the boundary condition adopted in Sec. \ref{secMWVD} corresponds to the extruding front; compare Fig. \ref{FigSch}(a) with the region illustrated by the dotted line in Fig. \ref{FigFr} (note that the sign of $s$ should be changed in Fig. \ref{FigSch}(a) for the comparison).
Therefore, we consider here only the extruding one. 
Actually, the region where $\dot{u}$ is almost zero is stable, and the front does not exist there.
Additionally, the friction stress has exceeded the maximum value and become almost zero at the front shown in Fig. \ref{FigFr}, which is consistent with the definition of the slip front in Sec. \ref{secMWVD}. 

Let us begin analytical treatment with Eq. ($\ref{eqeom4}$). If we substitute $\exp[-i (k x - \omega t) ]$ into Eq. (\ref{eqeom4}), the dispersion relation is easily shown to be
\begin{equation}
-\omega^2 = - k^2 - i \eta \omega k^2 + 2iab \omega.
\end{equation}
The real and imaginary parts of the dispersion relation are
\begin{equation}
(\eta \omega_i - 1)(k_r^2 - k_i^2) + 2 \eta \omega_r k_r k_i + (\omega_r^2 - \omega_i^2) - 2ab \omega_i = 0, \label{eqD1}
\end{equation}
\begin{equation}
2 k_r k_i (\eta \omega_i - 1)  - \eta \omega_r (k_r^2 - k_i^2) + 2 \omega_r \omega_i + 2ab \omega_r = 0, \label{eqD2}
\end{equation}
respectively. Differentiating Eqs. ($\ref{eqD1}$) and ($\ref{eqD2}$) with respect to $k_r$ and employing the growth and propagating stabilities [Eqs. ($\ref{eqGI2}$) and ($\ref{eqPI2}$)] yield
\begin{equation}
2 (\eta \omega_i - 1) k_r + 2 \eta k_i (c k_r + \omega_r ) + 2 \omega_r c = 0, \label{eqD1d1}
\end{equation}
\begin{equation}
2 (\eta \omega_i - 1) k_i - \eta c (k_r^2-k_i^2) - 2 \eta \omega_r k_r + 2 \omega_i c + 2abc =0, \label{eqD2d1}
\end{equation}
respectively. Moreover, differentiating Eqs. ($\ref{eqD1}$) and ($\ref{eqD2}$) with respect to $k_i$ and employing the growth and propagating stabilities [Eqs. ($\ref{eqGI1}$) and ($\ref{eqPI1}$)] give
\begin{equation}
\eta c (k_r^2-k_i^2) -2(\eta \omega_i - 1) k_i + 2 \eta \omega_r k_r -2 \omega_i c - 2abc =0, \label{eqD1d2}
\end{equation}
\begin{equation}
2 \eta \omega_i k_r + 2(\eta \omega_i - 1) k_r + 2 \eta \omega_r k_i + 2 \omega_r c =0, \label{eqD2d2}
\end{equation}
respectively. We have four unknown variants $k_r, k_i, \omega_r, \omega_r$, though there exist six equations ($\ref{eqD1}$)-($\ref{eqD2d2}$). However, because Eqs. ($\ref{eqD1d1}$) and ($\ref{eqD2d2}$), and Eqs.  ($\ref{eqD2d1}$) and ($\ref{eqD1d2}$) are exactly the same, respectively, the independent equations are ($\ref{eqD1}$), ($\ref{eqD2}$), ($\ref{eqD1d1}$) and ($\ref{eqD2d1}$). In addition, Eq. ($\ref{eqD2}$) and Eq. ($\ref{eqD2d1}$) give $2 \eta \omega_r k_r^2=0$, which concludes $k_r=0$ or $\omega_r=0$. Moreover, if $k_r=0$ or $\omega_r=0$, we can conclude that $k_r=\omega_r=0$ based on the propagation stability ($\ref{eqPI2}$).

Employing $k_r = \omega_r =0$, the left hand sides of Eqs. ($\ref{eqD2}$) and ($\ref{eqD1d1}$) are identically zero. We also have
\begin{equation}
(- \eta \omega_i + 1) k_i^2 - \omega_i^2 - 2ab \omega_i=0, \label{eqI1}
\end{equation}
\begin{equation}
2 (\eta \omega_i - 1) k_i + \eta c k_i^2 +2 \omega_i c + 2abc =0, \label{eqI2}
\end{equation}
from Eqs. ($\ref{eqD1}$) and ($\ref{eqD2d1}$), respectively. In addition, dividing Eq. ($\ref{eqI1}$) by $k_i$ and employing the relationship $\omega_i/k_i = c$, we obtain an equation
\begin{equation}
(- \eta \omega_i +1) k_i - \omega_i c - 2abc = 0. \label{eqI3}
\end{equation}
We will obtain $k_i, \ \omega_i, \ c$ from Eqs. ($\ref{eqI1}$), ($\ref{eqI2}$) and ($\ref{eqI3}$). 
First, Eqs. ($\ref{eqI2}$) and ($\ref{eqI3}$) give
\begin{equation}
k_i^2 = \frac{2ab}{\eta}. \label{eqks1}
\end{equation}
We have $k_i = \sqrt{2ab/\eta}$ from this equation since $k_i$ is assumed to be positive. With this result and Eq. ($\ref{eqI1}$), we can see that $\omega_i$ obeys the equation
\begin{equation}
\omega_i^2 + 4ab \omega_i -\frac{2ab}{\eta} = 0, \label{eqo2}
\end{equation}
which gives the solution
\begin{equation}
\omega_i = - 2ab \pm \sqrt{4a^2 b^2 + \frac{2ab}{\eta} }. \label{eqOmega}
\end{equation}
We should select the plus sign and write the solution
\begin{equation}
\omega_i = \sqrt{4a^2 b^2 + \frac{2ab}{\eta}} - 2ab,
\end{equation}
since $\omega_i$ is assumed to be positive. This equation together with Eq. ($\ref{eqks1}$) give the SFP velocity $c$ in the form
\begin{eqnarray}
c &=& \frac{\omega_i}{k_i} = \left( \sqrt{ 4a^2b^2 + \frac{2ab}{\eta}} - 2ab \right) \sqrt{\frac{\eta}{2ab}} \nonumber \\
&=& \sqrt{1 + 2ab \eta} - \sqrt{2ab \eta}. \label{eqc}
\end{eqnarray}
We have $v^{\mathrm{ex}}=\sqrt{1+2ab \eta}-\sqrt{2ab \eta}$, which is smaller than the elastic wave velocity and exactly the same as $v_{c-}$.

The slip front treated in Sec. \ref{secMWVD} was the extruding front, which is consistent with the statement that $v^{\mathrm{ex}}=v_{c-}$. On the other hand, as mentioned in Sec. \ref{secExact2}, $v_{c+}$ describes the SFP velocity with the situation where the slip velocity is initially $2b$ in the whole region and we arrest the slip at $x \to \infty$. Note here that if the propagation direction is leftward in Fig. \ref{FigSch}(a), we can confirm that the slip front with $v^{\mathrm{in}}$ will emerge [see also Fig. \ref{FigSch}(b)]. Additionally, if we substitute $\exp [i (k x -\omega t)]$ into Eq. (\ref{eqeom4}), we easily obtain $v^{\mathrm{in}}=\sqrt{1+2ab \eta}+\sqrt{2ab \eta}$, which concludes $v^{\mathrm{in}}=v_{c+}$.

Another important conclusion here is that only the linearized form of the friction law is relevant to the SFP velocity. 
Only the requirement for the treatment here is that the friction law can be expanded by $\dot{u}$ around the point where the friction stress vanishes. 
These conclusions are the same as obtained in Sec. \ref{secExact1}. We can suggest that detailed information about the friction stress may not be required for future studies on SFP.

\subsection{Numerical calculations} \label{secNum2}

\begin{figure}[tbp]
\centering
\includegraphics[width=8.5cm]{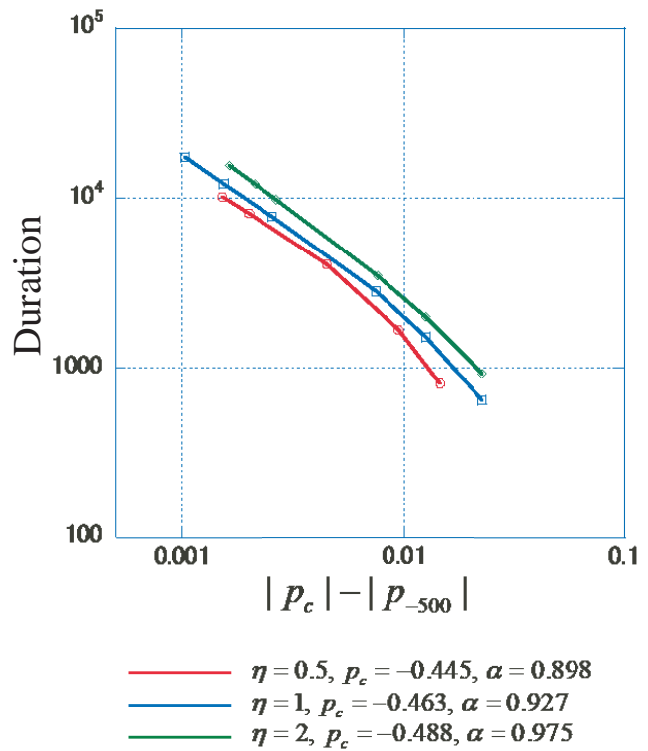}
\caption{The slip duration in terms of $|p_{-500}-p_c|$. The red, blue and green curves describe the cases $(\eta, p_c) = (0.5, -0.445), (1, -0.463)$ and  $(2,-0.488)$, respectively. 
The values of $p_c$ and $\alpha$ are derived based on the least-square method by approximating the curves by $|p_c-p_{-500}|^{\alpha}$.}
\label{FigSD}
\end{figure}

Numerical calculations based on the model described in Sec. \ref{subAna} are performed in this section. 
We assume here that the side-loading point $x=-500$ is regarded as the point $x \to -\infty$, as done in Sec. \ref{secMWVD}, so that $p|_{x_3 \to -\infty}$ in the analytical discussion above can be replaced with $p_{-500}$. 
Note that even though we assumed $0<\dot{u}<2b$ in Sec. \ref{secExact2}, the solution allowing $\dot{u}>2b$ can emerge.
Therefore, difference between the analytical solution for the SFP velocity obtained in Secs. \ref{secExact2} and \ref{secLMSH} and the numerical solution may arise, as discussed later in the present section. 
We first obtain the condition on $p_{-500}$ for the steady SFP by considering the slip duration as a function of $|p_{-500}|$ (Fig. \ref{FigSD}; values of $(a, b)=(0.1, 0.2)$ are fixed in the figure and henceforth). 
The slip duration is defined as the time when the slip velocity at all of the points on the slip plane turns to be smaller than $0.1 |p_{-500}|$ from the onset of the side-loading. 
We confirmed the value $0.1$ does not affect the result.
If the slip duration is finite, the steady SFP does not exist. 
Figure \ref{FigSD} implies that the slip duration diverges with power law  $(|p_c|-|p_{-500}|)^{-\alpha}$ in the region $|p_{-500}|<|p_c|$, where $p_c (<0)$ and $\alpha (>0)$ are the critical strain and the critical exponent, respectively. 
This result indicates that the steady SFP exists in the region $|p_{-500}|>|p_c|$. 
It is to be noted that the value of $p_c$ depends on $\eta$.

\begin{figure}[b]
\centering
\includegraphics[width=8.5cm]{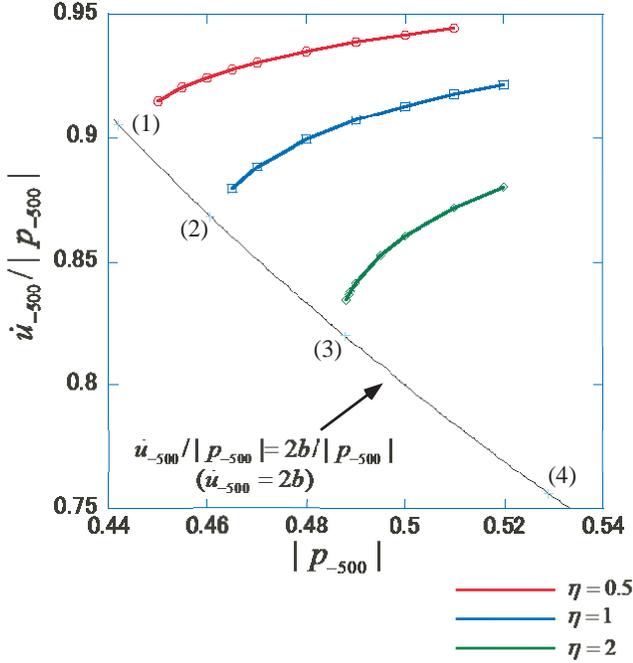}
\caption{The value of $\dot{u}_{-500}/|p_{-500}|$. The red, blue and green curves illustrates the cases $\eta=0.5, \ 1$ and $2$, respectively. 
The thin black solid line indicates the relationship $\dot{u}_{-500}/|p_{-500}|=2b/|p_{-500}|$. 
The light blue crosses stand for the points $(2b/(\sqrt{1+ab \eta}-\sqrt{ab \eta}), \sqrt{1+ab \eta}-\sqrt{ab \eta})$ with $\eta=0.5 \ (1), \ 1 \ (2), \ 2 \ (3)$ and $4 \ (4)$. If we consider the points $(2b/(\sqrt{1+2ab \eta}-\sqrt{2ab \eta}), \sqrt{1+2ab \eta}-\sqrt{2ab \eta})$, the four points correspond to $\eta=0.25 \ (1), \ 0.5 \ (2), \ 1 \ (3)$ and $2 \ (4)$, respectively. 
The coordinate values of points (1)-(4) are (1) (0.442, 0.905), (2) (0.461, 0.869), (3) (0.488, 0.820), and (4) (0.529, 0.756).
}
\label{FigVL}
\end{figure}

We now investigate the SFP velocity in detail in the region $|p_{-500}|>|p_c|$. 
We show the value of $\dot{u}_{-500}/|p_{-500}|$ in the steady state in Fig. \ref{FigVL}, where $\dot{u}_{-500} \equiv \lim_{t \to \infty} \dot{u}(-500,t)$. 
This value describes the SFP velocity in the steady state because the relationship $\dot{u}=\partial u/ \partial t =-v \partial u/ \partial x=-v p$ is satisfied in the steady state moving with the constant velocity $v$. 
%
The figure shows that the SFP velocity is less than the SFP velocity in the absence of the friction, which is unity here.
Moreover, the minimal value also exists for the SFP velocity.
The extrapolated curves of $\dot{u}_{-500}/|p_{-500}|$ cross the curve $2b/|p_{-500}|$ near the point $(2b/(\sqrt{1+ab \eta}-\sqrt{ab \eta}), \sqrt{1+ab \eta}-\sqrt{ab \eta})$. 
This concludes that $\sqrt{1+ab \eta}-\sqrt{ab \eta}$ gives the minimal value for the SFP velocity.
It should also be noted that $p_c$ is approximately given by $-2b/(\sqrt{1+ab \eta}-\sqrt{ab \eta})$, not $-2b/(\sqrt{1+2ab \eta}-\sqrt{2ab \eta})=-2b/v_{c-}$.

\begin{figure}[tbp]
\centering
\includegraphics[width=4.5cm]{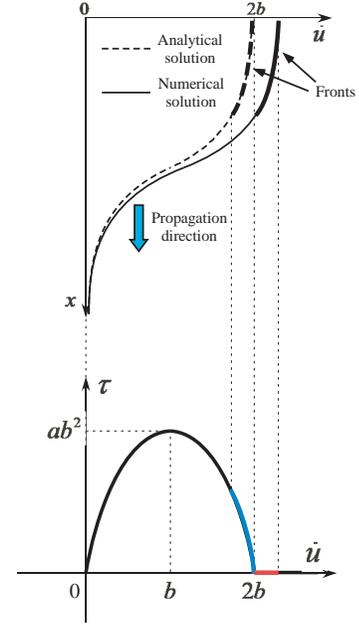}
\caption{Comparison between the analytical and numerical solutions. 
In the upper part, the slip velocity profiles for the analytical (dashed) and the numerical (solid) solutions are illustrated. 
In the bottom part, the constitutive law between the friction stress and the slip velocity is shown (see also Fig. $\ref{FigCL}$). 
}
\label{FigGeo}
\end{figure}

\begin{figure}[tbp]
\centering
\includegraphics[width=7.5cm]{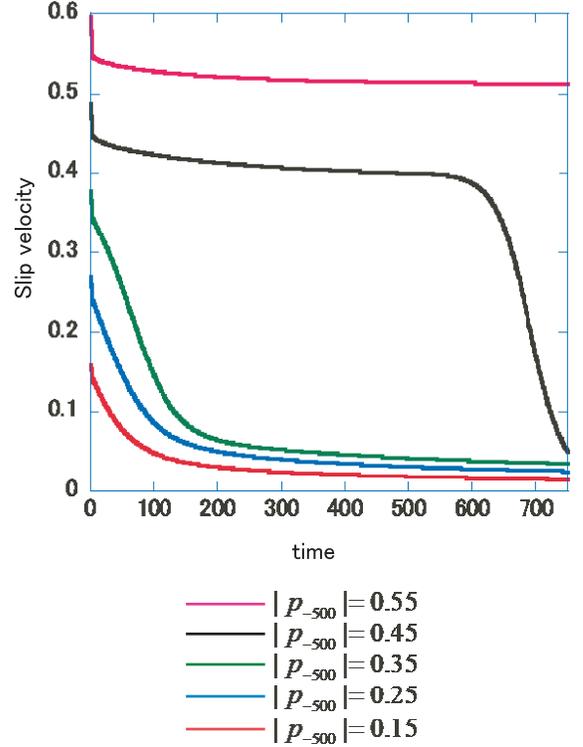}
\caption{Temporal changes in the slip velocity at the loading point $x=-500$. 
The value $\eta=1$ is employed. 
The parameter set $(a, b)$ is (0.1,0.2). The cases $|p_{-500}|=0.15, 0.25, 0.35, 0.45$ and $0.55$ are calculated.}
\label{FigDec2}
\end{figure}

\begin{figure}[tbp]
\centering
\includegraphics[width=8.5cm]{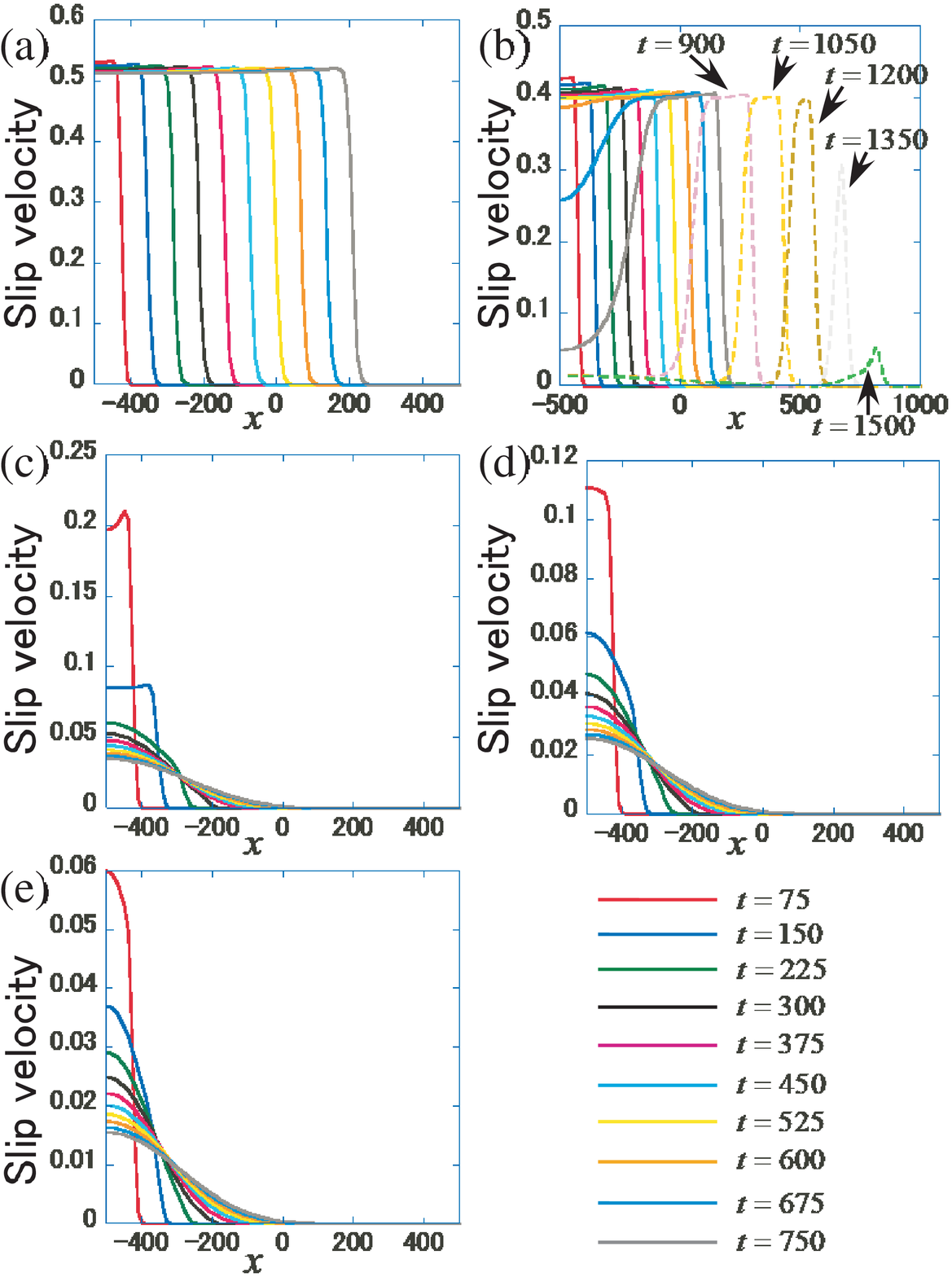}
\caption{Spatiotemporal changes in the slip velocity profile. 
The value $\eta=1$ is employed. 
The five cases $p_{-500}=$ (a) $0.55$, (b) $0.45$, (c) $0.35$, (d) $0.25$, and (e) $0.15$ are calculated.  
Difference in line colors describes that of the time. Note that the calculation is performed until $t=1500$ only for (b).}
\label{FigSTV2}
\end{figure}

To understand these results, note that slip front is defined at the point where the friction stress is almost negligible near $\dot{u}=2b$. 
For the analytical solution derived in Sec. \ref{secExact2}, the condition $\dot{u}<2b$ is assumed at all of the points on the slip plane and the non-zero friction stress works at the slip front.
See the blue curve in Fig. $\ref{FigGeo}$. 
However, note that the slip velocity can exceed $2b$ near the slip front when $|p_{-500}|>|p_c|$ (note the curves illustrating $\dot{u}_{-500}/|p_{-500}|$ are on the upside of the curve $2b/|p_{-500}|$ in Fig. \ref{FigVL}) for numerical solutions. 
The analytical solution for the SFP velocity cannot be applied directly because the friction stress is no longer on the parabola form as a function of the slip velocity when $\dot{u}>2b$.
See the red line on the $\dot{u}$ axis shown in Fig. $\ref{FigGeo}$. 
The friction stress around the slip front for the numerical calculation is concluded to be always smaller than that for the analytical solution, so that the SFP velocity for the former case is larger than that for the latter case. 
However, we should emphasize that the friction stress actually works just ahead of slip front, so that the propagation velocity does not approach the value of unity. 
The SFP velocity is concluded to be larger than the analytical solution $v_{c-}$ and smaller than unity, while we cannot obtain its exact analytical solution. 
We can also confirm the steady SFP from the case $|p_{-500}|=0.55$ shown in Figs. \ref{FigDec2} and \ref{FigSTV2}.

The steady SFP does not exist for $|p_{-500}|<|p_c|$ [Figs. \ref{FigSTV2}(b-e)]. 
For example, if $|p_{-500}|=0.45$ and $|p_c| \sim 2b/(\sqrt{1+ab \eta}-\sqrt{ab \eta}) =0.461$ as the case shown in Fig. $\ref{FigSTV2}$ (b), the slip velocity decreases and approaches zero with increasing time, and non-steady pulse-like slip behavior appears. 
For the cases $|p_{-500}|<|p_c|$, the stable state is given by that where uniformly $\dot{u}$ and $p$ approach zero and $p_{-500}$, respectively, as shown in Sec. \ref{secMWVD}.

We emphasize that the critical SFP velocity is estimated here and it is smaller than that without the viscosity. The value $v_{c-}=\sqrt{1 +2ab \eta}-\sqrt{2ab \eta}$ gives rough (not exact) estimation of the smallest SFP velocity, and setting $\eta$ as $\eta/2$ for $v_{c-}$ seems to be better estimation.

\section{DISCUSSION AND CONCLUSIONS} \label{secDisCon}

The 1D model is assumed for treating the onset of the macroscopic slip. The local friction law is assumed to be the quadratic form of the slip velocity with no static friction. The lower limit of the side-loading strain at the loading point to slide the block is obtained, and it corresponds to the macroscopic static friction stress. We can interpret that the macroscopic static friction stress emerges spontaneously. The SFP velocity is found to be given by the elastic wave velocity (unity in the nondimensionalized system). Even if we introduce the viscosity into such a framework, the macroscopic static friction stress spontaneously emerges again, while the value depends on the viscosity. The estimations for the SFP velocities are given for the with-viscosity case based on the analytical treatment. One of them corresponds to the boundary condition used here, and it is smaller than that for the without-viscosity case. The linear marginal stability hypothesis (LMSH) gives the same solutions for the SFP velocities, even though the governing equation is linearized. This statement leads to the conclusion that the detail of the dependence of the friction stress on the slip velocity does not affect the front velocity. The only assumption required for using LMSH is that the friction stress vanishes above a certain value of the slip velocity. Note that the analytical solution for the SFP velocity associated with the boundary condition here is smaller than that numerically obtained, since the friction stress deviates from the parabola form as a function of the slip velocity. Nonetheless, the analytical solution gives rough estimate for the SFP velocity.

The framework constructed in this paper can be extended to other systems. For example, the slip front velocity in \cite{Ots} can be evaluated based on the result here, while they did not calculate it. Although they assumed the friction stress decreasing linearly with the slip velocity, we can roughly estimate the parameter values $a$ and $b$ by approximating the linear decrease by a parabola; we consider that $ab^2 \sim (\mu_S-\mu_K) \tilde{P}_{\mathrm{ext}}$ and $2b \sim \tilde{v}_c$ with their notation. With these assumptions, the parameters are given by $a \sim 10^4$ and $b \sim 10^{-4}$ for the numerical calculation shown in Fig. 1 in their paper ($\mu_S=0.38, \mu_K=0.1, \tilde{P}_{\mathrm{ext}}=0.003$ and $\tilde{v}_c=3.4 \times 10^{-4}$). In addition, the parameter $\eta$ is unity there. The value $\sqrt{1+ab \eta}-\sqrt{ab \eta}$ is estimated as 0.41 with these parameters, and this can be a rough estimation for the front velocity observed in their Fig. 1. 

We then give some implications for the creep motion observed when the side-loading stress is smaller than the macroscopic static friction stress. Such a creep motion may be related with precursors preceding macroscopic slip because its propagation will be arrested easily by negligibly small perturbation of the stress on the plane, and it leaves stress profile disturbance there. Such stress disturbance may induce macroscopic slip with the side-loading stress smaller than the critical value. Laboratory experiments can be explained with this viewpoint \cite{Ben}. However, as mentioned in Sec. \ref{secI}, note that the front propagation velocity can be smaller or larger than the elastic wave velocity. The smaller one can be modeled in the present framework, and treating the larger one has potency of a future work. 

Finally, we give seismological implications from the results. For example, the pulse-like slip shown in Fig. \ref{FigSTV2}(b) may explain slip behaviors observed for usual earthquakes (e.g., \cite{Wal}). Although the pulse-like slip has been explained in terms of, e.g., the dilatancy effect of the fault rocks \cite{Suz08}, the nonlinear friction law itself can generate such a slip. Furthermore, note that if $|p_{-\infty}| \ll 2b$, the emergent slip velocity is negligibly small. Slow earthquakes may be related with this behavior. We have succeeded in dynamic modeling of slow earthquakes in terms of the friction law nonlinearly depending on the slip velocity.

\acknowledgments
This research was supported by JSPS KAKENHI Grant Number JP26400403 and JP16K17795 in Scientific Research on Innovative Areas ``Science of Slow Earthquakes''.
T. S. is also supported by JSPS KAKENHI Grant Number JP16H06478.
H. M. is also supported by JSPS KAKENHI Grant Number JP17K05586.
This study was supported by the Earthquake Research Institute cooperative research program.


\appendix

\renewcommand{\theequation}{A\arabic{equation}}
\setcounter{equation}{0}

\section{EXACT SOLUTION FOR THE STEADY STATE BASED ON EQ. ($\ref{eqGov}$)} \label{secAA}

We consider here the steady state solution of Eq. (\ref{eqGov}) propagating with the constant SFP velocity $v$, which means $u(x,t)$ has a form $u(x-vt)$. 
We treat the solution $0<\dot{u} <2b$ in the whole plane, so $H(\dot{u})-H(\dot{u}-2b)=1$. Under these conditions, Eq. (\ref{eqGov}) leads to
\begin{equation}
(v^2-1) u'' = a v^2 u'^2 +2abvu' = avu' ( vu' +2b). \label{eqss}
\end{equation}
The prime describes the differentiation with respect to $x_1=x-vt$. 
Equation (\ref{eqss}) is expressed as
\begin{equation}
(v^2-1) \frac{dp}{dx_1} = avp(vp+2b), \label{eqpp}
\end{equation}
where $p(x_1)=du(x_1)/dx_1$ is the strain. 
As noted in the text, the boundary conditions are given by $p(x_1=-\infty)=p_{-\infty}$ and $p(x_1=\infty)=0$. 
We can integrate Eq. (\ref{eqpp}) to obtain the steady propagating solution,
\begin{equation}
\frac{p}{vp+2b} = C_1 e^{-\gamma x_1}, \label{eqp1}
\end{equation}
where $C_1$ is a constant and $\gamma = 2abv / (1-v^2)$. 

The condition $p(\infty)=0$ and Eq. (\ref{eqp1}) require $\gamma>0$, i.e., $v$ must be smaller than the elastic wave velocity, unity. 
The condition $p(-\infty)=p_{-\infty}$ and Eq. (\ref{eqp1}) lead to $p_{-\infty} =-2b/v$, which determines $v$. 
Here we note $p(x_1=0)=p_0<0$ and have solutions for $p$ and $u$,
\begin{equation}
p(x_1)=\frac{du}{dx_1}=-\frac{2b}{v} \frac{vp_0 e^{-\gamma x_1}}{vp_0 e^{-\gamma x_1} -vp_0-2b}, \label{eqp}
\end{equation}
\begin{equation}
u(x_1) = \frac{2b}{\gamma v} \ln \left( 1 - \frac{vp_0}{2b+vp_0} e^{-\gamma x_1} \right), \label{eqan-uA}
\end{equation}
respectively. 
Note that $p$ must satisfy the relationship $-2b/v < p < 0$ because the left hand side of Eq. (\ref{eqp1}) should not change its sign. Therefore, the slip velocity given by $\dot{u}=-v du/dx_1 = -v p$ always satisfies $0<\dot{u}<2b$, which is consistent with the assumption noted above. 
Additionally, the slip velocity profile is given by differentiating Eq. (\ref{eqan-uA}) with respect to $t$.

\renewcommand{\theequation}{B\arabic{equation}}
\setcounter{equation}{0}

\section{INTRODUCTION OF LINEAR MARGINAL STABILITY HYPOTHESIS} \label{secAB}

We give detailed explanations for the growth and propagating instabilities here. First, the growth stability gives the condition
\begin{equation}
\frac{\partial \omega_i}{\partial k_r}=0,  \label{eqABGI1}
\end{equation}
where $\omega_i$ and $k_r$ are the imaginary part of the frequency and the real part of the wave number, respectively. To understand the physical meaning of this relationship, let us assume that the wave number has small disturbance, $k=k_0 + \Delta k \ ( \Delta k \in \mathbb{R})$, where $k_0$ is a constant complex number. Let us take into account of this assumption and the following relationship:
\begin{eqnarray}
|\exp (\pm i(kx-\omega t))| &=& |\exp(\pm i ((k_r + i k_i)x - (\omega_r + i \omega_i)t))| \nonumber \\
&=& |\exp(\mp (k_i x - \omega_i t)) \exp(\pm i (k_r x - \omega_r t))| \nonumber \\
&\sim& \exp(\mp(k_i x - \omega_{i0} t) \exp(\pm \frac{\partial \omega_i}{\partial k_r} \Delta k \cdot t), \label{eqABAm}
\end{eqnarray}
where $k_i$ and $\omega_r$ are the imaginary part of the wave number and the real part of the frequency, respectively, and $k_{i0}$ and $\omega_{i0}$ are imaginary parts of $k_0$ and $\omega(k=k_0)$, respectively. We can conclude that if $\partial \omega_i / \partial k_r \ne 0$, we can select $\Delta k$ satisfying $(\partial \omega_i / \partial k_r) \Delta k > 0$, which induces the exponential increase in the amplitude of the disturbance. Such increase does not generate the steady front propagation, so that Eq. (\ref{eqABGI1}) assures stability of the growth of disturbance.

Second, the propagating stability gives the relationship
\begin{equation}
\frac{\omega_i}{k_i} = \frac{\partial \omega_i}{\partial k_i}. \label{eqABPI1}
\end{equation}
Equation (\ref{eqABPI1}) clearly shows that the phase velocity $c_p$ (the left hand side) equals to the group velocity $c_g$ (the right hand side). Since the disturbance propagates with the group velocity, this condition describes that the disturbance and the front propagate with the same velocity. However, it is important that for the stability, the relationship $c_p \ge c_g$ is sufficient because the disturbance is overtaken by the front with such a relationship. Nonetheless, it is mathematically shown that the relationship (\ref{eqPI1}) is satisfied for spontaneous front propagation (e.g., \cite{Saa1}). We refer to this velocity as $c \ (=c_p=c_g)$.

We need two more requirements to determine all variants. If we assume $\omega$ is a regular function of $k$, we have the requirements
\begin{equation}
\frac{\partial \omega_r}{\partial k_i} =0, \label{eqABGI2}
\end{equation}
\begin{equation}
\frac{\partial \omega_r}{\partial k_r} =\frac{\omega_r}{k_r} = c, \label{eqABPI2}
\end{equation}
from the Cauchy-Riemann relationship, the growth and propagating stability conditions, and the requirement from the viewpoint of the steady front propagation (the front and the disturbance should propagate with the same velocity also for the real part of $\exp[\pm (kx-\omega t)]$, so that $\partial \omega_r/ \partial k_r = \omega_r / k_r$ must be satisfied).




\end{document}